\documentclass[english,aps,preprint]{revtex4}
\usepackage[T1]{fontenc}
\usepackage[latin9]{inputenc}
\usepackage{graphicx}
\usepackage{esint}
\usepackage[labelfont=bf,labelsep=period,justification=raggedright]{caption}

\makeatletter

\newcommand{\noun}[1]{\textsc{#1}}

\newcommand{\qts}[1]{``#1''}

\@ifundefined{textcolor}{}
{%
 \definecolor{BLACK}{gray}{0}
 \definecolor{WHITE}{gray}{1}
 \definecolor{RED}{rgb}{1,0,0}
 \definecolor{GREEN}{rgb}{0,1,0}
 \definecolor{BLUE}{rgb}{0,0,1}
 \definecolor{CYAN}{cmyk}{1,0,0,0}
 \definecolor{MAGENTA}{cmyk}{0,1,0,0}
 \definecolor{YELLOW}{cmyk}{0,0,1,0}
}

\makeatother

\usepackage{babel}
\begin{document}

\title{Ergodicity, Hidden Bias and the Growth Rate Gain}

\author{Nash D. Rochman}

\affiliation{Johns Hopkins University, Department of Chemical and Biomolecular
Engineering}

\author{Dan M. Popescu}

\affiliation{Johns Hopkins University, Department of Applied Mathematics and Statistics}

\author{Sean X. Sun}
\email{ssun@jhu.edu}

\affiliation{Johns Hopkins University, Departments of Mechanical Engineering and
Biomedical Engineering}
\begin{abstract}
Many single-cell observables are highly heterogeneous. A part of this heterogeneity stems from age-related phenomena: the fact that there is a nonuniform distribution of cells with different ages.
This has led to a renewed interest in analytic methodologies including use
of the \qts{von Foerster equation} for predicting population growth
and cell age distributions. Here we discuss how some of the most popular
implementations of this machinery assume a strong condition on the
ergodicity of the cell cycle duration ensemble. We show that one common
definition for the term ergodicity, ``a single individual observed over many generations
recapitulates the behavior of the entire ensemble'' is implied by the other, ``the probability
of observing any state is conserved across time and over all individuals'' in an ensemble
with a fixed number of individuals but that this is not true
when the ensemble is growing. We further explore the impact 
of generational correlations between cell cycle durations on the population growth rate. Finally,
we explore the \qts{growth rate gain} - the phenomenon that variations in
the cell cycle duration lead to an improved population-level growth rate - in
this context. We highlight that, fundamentally, this effect is due
to asymmetric division.
\end{abstract}
\maketitle

\section{Introduction}

In most cell biology experiments, measurements are made on cells during exponential growth, and the results are often very heterogeneous from cell to cell. This has led to the prevalent use of $p$-values and other sophisticated statistical measures to determine if two populations are likely to be different. There are several fundamental reasons for this observed variability\cite{metz2014dynamics}.
ranging from stochastic protein synthesis rates to the natural selection of diverse, adaptable populations\cite{rochman2016grow,bodi2017phenotypic}. One of the most important reasons is that cells are at different stages of the cell cycle, and therefore display different cycle-dependent protein expression levels. For cells replicating mitotically, there are more younger cells than older cells present in a random selection of individuals; therefore, the age (as measured from the moment of completion of mitosis) distribution of the cells determines, to a large extent, the statistical distribution of observables. The basic formalism for modeling the age distribution of a population was 
developed by von Foerster \cite{von1960doomsday}. Starting from a known cell cycle duration (the time between cell birth and cell division) distribution, the expected age distribution can be explicitly computed \cite{powell1956growth,stukalin2013age}; however, the von Foerster approach assumes a strong condition for the ergodicity of the cell population. In particular, this formalism implicitly assumes that the cell cycle durations from generation to generation are independently sampled from the same distribution across all individuals. Recent experimental work has shown measurable correlations in cell cycle duration contradicting this assumption for both bacterial \cite{wang2010robust,hashimoto2016noise} and yeast populations \cite{cerulus2016noise}. These measurements have been made in a variety of settings including microfluidic devices where bacteria are maintained at constant density and media conditions in channels of around ten to twenty cells\cite{wang2010robust} and up to one hundred cells\cite{hashimoto2016noise}, as well as agar sandwiches for yeast where density remains low throughout the experiment, though it is not explicitly controlled\cite{cerulus2016noise}. Moreover, it has been shown that higher variability in the cell cycle duration distribution leads to higher population-level growth rates \cite{cerulus2016noise,hashimoto2016noise}, a phenomenon labelled the \qts{growth rate gain}. In this paper, we explicitly examine the relationships between generational correlations within the cell cycle duration distribution, width of the cell cycle duration distribution, and the population-level growth rate. We show that the age distribution obtained from observing the cell cycle duration distribution differs if observations are made of an ensemble of fixed size (e.g. within a microfluidic device where only one of two daughter cells remains in the ensemble under observation after cell division or randomly selected from a dividing population. We discuss how the width of the cell cycle duration distribution affects the generational cell cycle duration correlations predicted using the von Foerster formalism and highlight that, fundamentally, the \qts{growth rate gain} in this context is due to asymmetric division.

Consider the following conservation of population where $n(a,t)$
is the population per unit age at time t, $n(a+\Delta a,t+\Delta t)\Delta a=n(a,t)\Delta a-\lambda(a)n(a,t)\Delta a\Delta t$.
This states the population of cells at age $a+\Delta a$ and time
$t+\Delta t,$ is equal to the population of cells at age $a$ and
time $t$ after removing those that have left the ensemble (due to
mitosis, etc.). The loss rate, in units of $1/\left[t\right]$ is
notated as $\lambda(a)$. For the purposes of this paper, we may consider
\qts{chronological age}, $da/dt=1$, though in the case where the
\qts{age} of interest is cell phase or another non-chronological age,
this introduces some added complexity (please see\cite{rubinow1968maturity} equations 1 and 2).
Stated another way, in the case of chronological age, a cell is $a$-minutes older after
$a$-minutes of lab-frame time has passed. While this may be straight-forward,
it may not always be useful. For example, if you want
to compare a fast growing cell with a slow growing cell, both
will have the same chronological aging rate $da/dt=1$; however,
the fast growing cell will progress through the cell cycle
much faster. To capture this phenomenon, we may want to examine
the cell cycle "phase" rather than the chronological age. In the simpler case, $da/dt=1$,
we find the relatively simple form:
\begin{equation}
\label{eq:vanFoerster}
\frac{\partial n(a,t)}{\partial t}+\frac{\partial n(a,t)}{\partial a}=-\lambda(a)n(a,t)
\end{equation}
To solve this equation, we can examine the stage when the cell  is undergoing
exponential growth, $n(a,t)=N_{0}\exp(bt)g(a)$ where $b$
is the population growth rate and $g(a)$ is the steady state age distribution.
Using this assumed form, we may solve for $g(a)=g(0)\exp\left[-ba-\int_{0}^{a}\lambda(a')da'\right]$.
To move forward, we need some information about $\lambda(a)$. We
are focusing on the case where the cell multiplies through mitosis.
In this case, when the death rate is negligible, $\lambda(a)$ is just the division rate, and can
be obtained directly from the cell cycle duration probability density function, $w\left(a\right)$. $w\left(a\right)$ has been explicitly measured in bacteria\cite{wang2010robust,stukalin2013age,rochman2016grow}, yeast\cite{cerulus2016noise}, and mammalian cells\cite{stukalin2013age}. We find
\begin{equation}
 \lambda(a)=\frac{w(a)}{1-\int_{0}^{a}w(a')da'};
\end{equation}
$\lambda(a)$ is the probability of undergoing mitosis per unit time at age $a$, and is 
the ratio of the population of cells observed to divide at age $a$
over the population of cells which have matured to age $a$ without
yet dividing. We note that this too is only valid in the chronological
case (this is clear from just the units: here $\lambda(a)$, which
as defined has units of $1/\left[t\right]$, takes the units of $w(a)$,
$1/\left[a\right]$).  $\lambda(a)$ also appears in the following boundary
condition: the number of cells at age zero and time $t$ is just twice
the number of cells that divided at time $t$: $n(0,t)=2\int_{0}^{\infty}\lambda(a)n(a,t)da$.
We may now rewrite this boundary condition in terms of $w\left(a\right)$
explicitly:
\begin{equation}
\label{eq:vFBoundary1}
1=2\int_{0}^{\infty}\exp\left(-ba\right)w(a)da
\end{equation}
and similarly rewrite the steady state age distribution to yield:
\begin{equation}
\label{eq:vFBoundary2}
g(a)=2b\exp\left(-ba\right)\left[\int_{a}^{\infty}w(a')da'\right]
\end{equation}
This framework (Eqs.~\ref{eq:vFBoundary1} and \ref{eq:vFBoundary2}) provides a way to calculate the population
growth rate, $b$, and the age distribution, $g\left(a\right)$, given
only the cell cycle duration distribution, $w\left(a\right)$; however,
it is built on some strong assumptions owing to the interpretation
of Eq.~\ref{eq:vanFoerster}. We have already discussed that this result is only valid
for the case of chronological aging, and noted that we assume cell
death is negligible. Beyond this, there is a third assumption which
leads to some important consequences we want to discuss. We rewrote
the division rate in terms of the cell cycle duration distribution,
$\lambda(a)=\frac{w(a)}{1-\int_{0}^{a}w(a')da'}$ and in doing so,
assumed that $w\left(a\right)$ is the same for every cell, at all
times. In other words, consider the transition probability (per unit time) between
successive cell cycle durations, $P\left(a_{n}\rightarrow a_{n+1}\right)$
where $a_{n}$ represents the cell cycle duration for the cell of
interest during generation $n$. In general, this function may have
some dependence on $a_{n}$, $a_{n+1}$, and even explicit dependence
on time; however, we have strictly assumed:
\begin{equation}
P\left(a_{n}\rightarrow a_{n+1}\right)=w\left(a_{n+1}\right)\equiv w\left(a\right)
\end{equation}
To see how this assumption impacts the growth rate, let us first take
a look at how $w\left(a\right)$ maps to $g\left(a\right)$ in exponentially
growing ensembles as well as ensembles containing a fixed number of
individuals where only one daughter cell remains in the ensemble under observation after
division.

\section{Age Distribution Properties}

\begin{figure}
\includegraphics[scale=0.43]{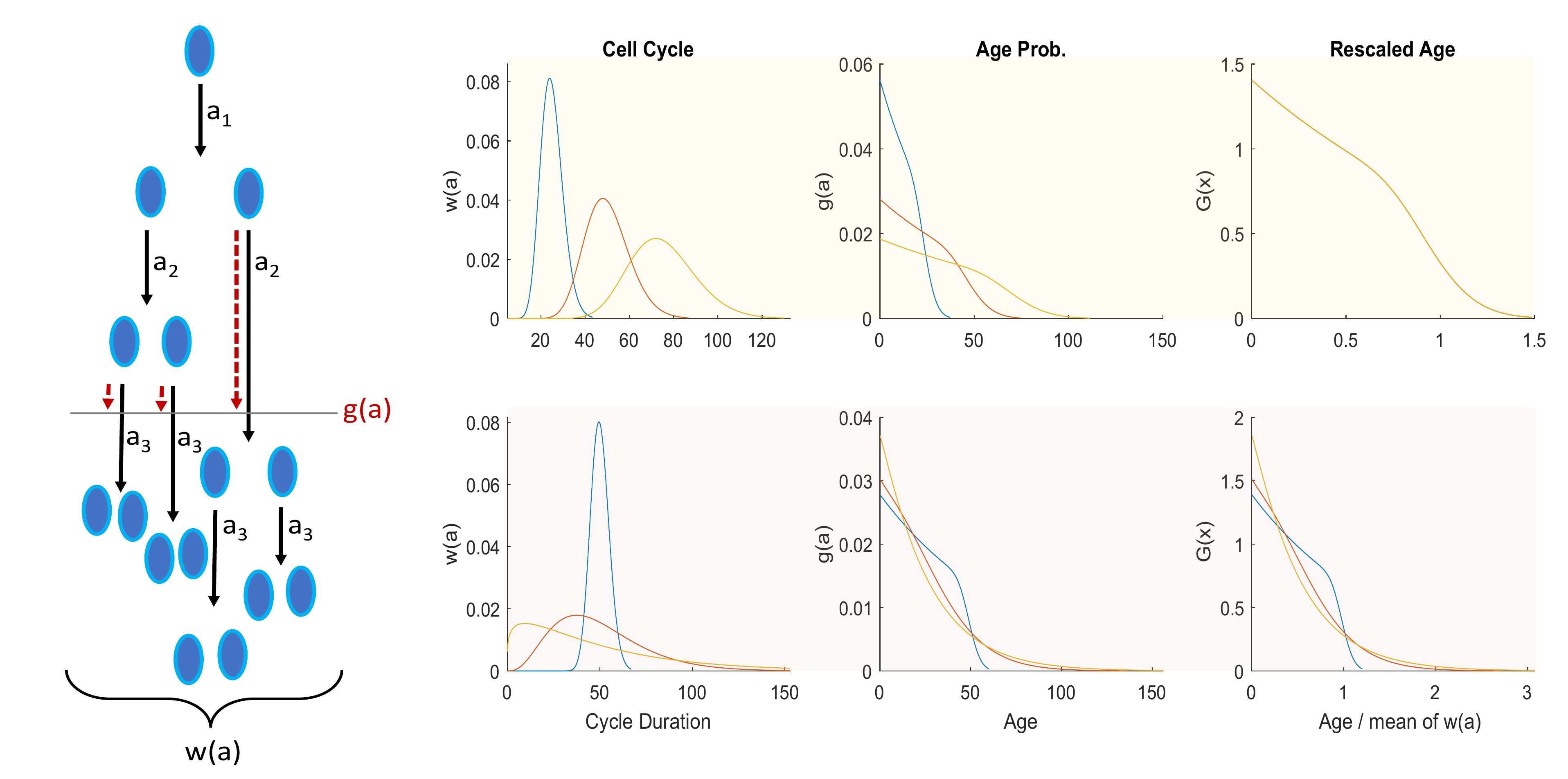}

\caption{Example $w\left(a\right)$ and their corresponding $g\left(a\right)$
and $G\left(x\right)$ . First Row: The mean is varied [25 (blue), 50 (red), 75 (yellow)]
while the $CV$ is kept constant $[0.2]$. Second Row: The CV is varied
[0.1 (blue), 0.5 (red), 0.9 (yellow)] while the mean is kept constant $[50]$. We see that
while the scaled age distribution is not conserved, the differences
between populations are still much smaller than that between the original
cell cycle duration distributions. The cartoon is meant to illustrate
how the age distribution, $g\left(a\right)$, is generated and that
it is weighted towards young cells.}
\label{fig:ExponentialEnsembleCartoon}
\end{figure}

First we will illustrate that, under the condition $P\left(a_{n}\rightarrow a_{n+1}\right)=w\left(a_{n+1}\right)\equiv w\left(a\right)$,
the age distribution is often \qts{mean-scalable}. Under
many circumstances, $w\left(a\right)$ belongs to a class of functions
which are mean-scalable\cite{stukalin2013age} (please note that\cite{stukalin2013age} does
not define the term "mean-scalable" but discusses measured $w\left(a\right)$ which 
satisfy the definition below). Let
us label the mean of $w\left(a\right)$ to be $\mu=\int a w(a) da$, and introduce the
variable $x=\frac{a}{\mu}$: we will say $w\left(a\right)$ is mean-scalable
when there exists a scaled function $\Omega\left(x\right)$ which
is conserved across all $\mu$ (and this function satisfies normalization):
\begin{equation}
\Omega\left(x\right)=\mu w\left(a\right);\,\int_{0}^{\infty}\Omega\left(x\right)dx=1
\end{equation}
Stated another way, suppose we have two functions $w_{1}\left(a\right)$ and $w_{2}\left(a\right)$
with mean values $\mu_{1}$ and $\mu_{2}$. Additionally, let us consider the functions
$\Omega_{1}\left(\frac{a}{\mu_{1}}\right)$ = $\mu_{1}w_{1}\left(a\right)$ and $\Omega_{2}\left(\frac{a}{\mu_{2}}\right)$ = $\mu_{2}w_{2}\left(a\right)$. If $\Omega_{1}\left(x\right)=\Omega_{2}\left(x\right)$ for all $x$,
then the family of functions consisting of $w_{1}\left(a\right)$ and $w_{2}\left(a\right)$ are mean-scalable. 
We can show that the mean-scalability of $w\left(a\right)$ confers
the same property to $g\left(a\right)$. Rewriting Eq.~\ref{eq:vFBoundary1} in terms
of $x$, $\Omega\left(x\right)$, and $B=\mu b$, the scaled bulk
growth rate, we find $1=2\int_{0}^{\infty}\exp\left(-Bx\right)\Omega\left(x\right)dx$.
Note that this uniquely defines $B$ given $\Omega\left(x\right)$
and that for fixed $B$, increasing $\mu$ decreases $b$. This inverse
relationship expresses that cells which take longer to divide result
in a slower growing ensemble. Similarly, we may now consider the function
$G\left(x\right)=\mu g\left(a\right)$ where the factor of $\mu$
is introduced again to maintain normalization. Rewriting Eq.~\ref{eq:vFBoundary2} yields
$G\left(x\right)=2B\exp\left(-Bx\right)\left[\int_{x}^{\infty}\Omega\left(x'\right)dx'\right]$.
We just saw that $B$ is completely determined by $\Omega\left(x\right)$
which implies that $G\left(x\right)$ is also completely determined
by $\Omega\left(x\right)$. In other words, whenever $w\left(a\right)$
is mean-scalable, $g\left(a\right)$ is also mean-scalable. This property
is useful because it has been observed that $w\left(a\right)$ is
often mean-scalable. More specifically, $w\left(a\right)$ is well
represented by a gamma distribution:
\begin{equation}\label{eq:gammadistribution}
w\left(a\right)=\frac{\beta^{\alpha}}{\Gamma\left(\alpha\right)}a^{\alpha-1}e^{-\beta a},a\geq0,\alpha>1,\beta>0
\end{equation}
where $\mu=\frac{\alpha}{\beta}$ and the coefficient of variation: $CV=\frac{1}{\sqrt{\alpha}}$,
across widely differing cell types. In this case $\Omega\left(x\right)=\frac{\alpha^{\alpha}}{\Gamma\left(\alpha\right)}x^{\alpha-1}e^{-\alpha x}$
and we see that the ensemble is no longer mean-scalable when the $CV$
(a function of $\alpha$) changes. We note, as shown in Fig.~\ref{fig:ExponentialEnsembleCartoon}, that even
when the $CV$ for the cell cycle duration distribution does vary
and $w\left(a\right)$ is not mean-scalable, the differences in the
resultant scaled age distributions, $G\left(x\right)$, are still
much smaller than the original cell cycle duration distributions.

As mentioned in the introduction, the von Foerster equation is usually
solved after making some important assumptions. One straightforward
assumption is that the species studied is undergoing exponential growth
$n(a,t)=N_{0}\exp(bt)g(a)$; but as clear as this may seem,
it is not valid for some experimentally relevant ensembles. A wide
variety of microfluidic devices and extracellular matrix patterns
are now available to study bacterial, yeast, and mammalian cell populations
at constant density, following a single cell for multiple cycles.
In these devices, only one of two daughter cells is maintained in
the ensemble after each division. This keeps the total number of cells
constant over time and modifies the age distribution. We could modify
the von Foerster equation to remove the time dependence and solve using the
same method as above; but below, we will utilize
a simpler framework. Let us take
a moment to look at the age distributions that result from these ensembles.

\begin{figure}
\includegraphics[scale=0.31]{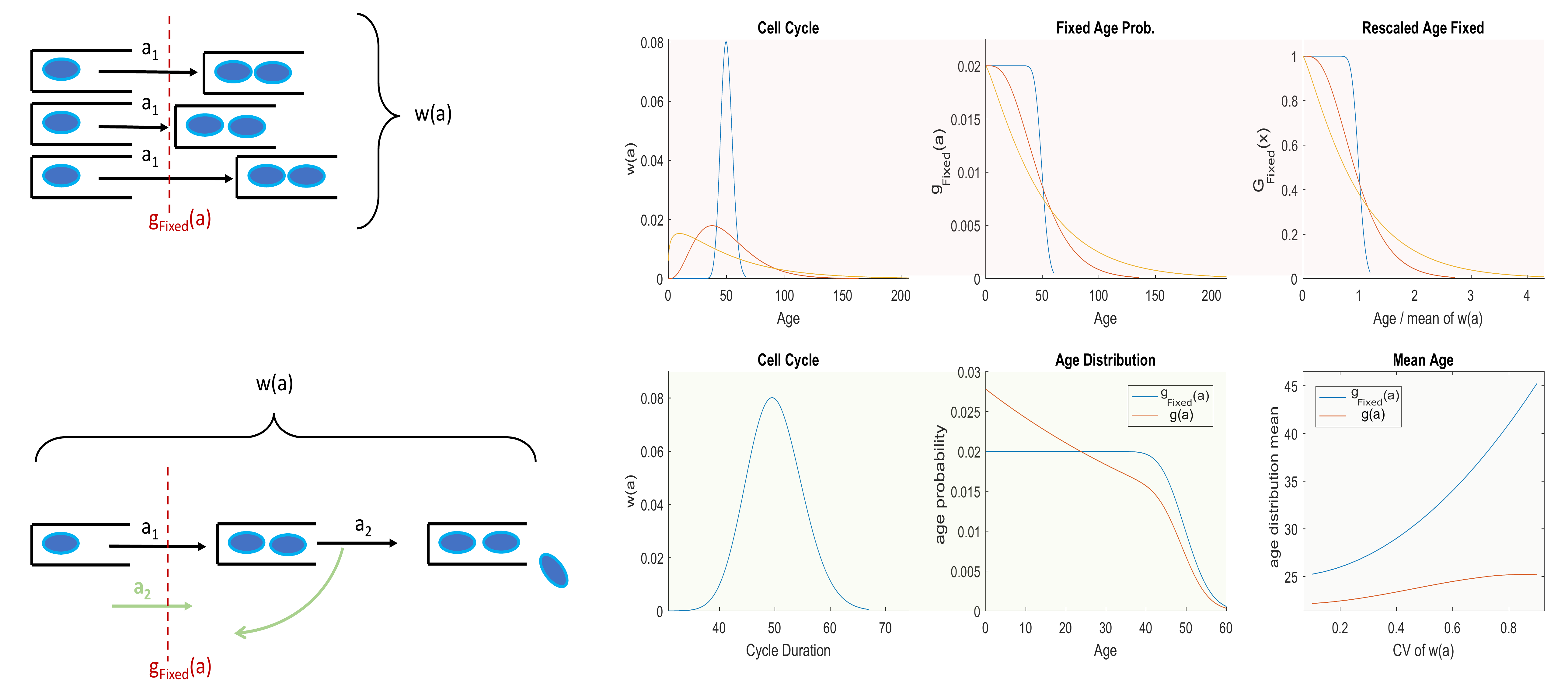}

\caption{Example $w\left(a\right)$ and their corresponding $g_{Fixed}\left(a\right)$
and $G_{Fixed}\left(x\right)$. First Row: the mean is kept constant
$[50]$ and the $CV$ is varied $[0.1,0.5,0.9]$. Second Row: a sample
cell cycle distribution $w\left(a\right)$, mean $[50]$ and $CV$
$[0.1]$; a comparison of $g\left(a\right)$ and $g_{Fixed}\left(a\right)$
for the sample $w\left(a\right)$; and the mean of $g\left(a\right)$
and $g_{Fixed}\left(a\right)$ resulting from cell cycle duration
distributions of fixed mean $[50]$ and varying $CV$ $[0.1,0.9]$.
Note that the inflection point displayed in the mean of $g\left(a\right)$,
in the bottom right plot, may be due to numerical error stemming from
highly skewed $g\left(a\right)$ when $w\left(a\right)$ has a large
$CV$ (the curve for $g_{Fixed}\left(a\right)$ is analytic). The
cartoon is meant to illustrate that following multiple generations
of one cell in/on a single channel/pattern is equivalent to observing
a single generation from multiple cells.}
\label{fig:FixedEnsembleCartoon}
\end{figure}

We begin again with the cell cycle duration distribution $w\left(a\right)$.
Unlike an exponentially growing ensemble, an ensemble with a fixed
number of individuals has very simple lineages: $a_{1}\rightarrow a_{2}\rightarrow...\rightarrow a_{N}$
where $a_{N}$ is the cell cycle duration of generation $N$. Since
we are still considering the condition $P\left(a_{n}\rightarrow a_{n+1}\right)=w\left(a_{n+1}\right)\equiv w\left(a\right)$,
we may also note that the behavior of each lineage recapitulates that
of the entire ensemble (this idea is discussed in detail in the following
section on ergodicity). Let us follow $N$ generations of
a single cell over some window of time $[t_{1},t_{2}]$ and observe just one of the two daughter cells after each division. The
probability of observing a cell cycle of duration exactly $a_{m}$ at an arbitrary time $t_{x}$ where $t_{1}<t_{x}<t_{2}$
is simply the duration of the cycle multiplied by the number of times it occurred.
$P_{obs}\left(a_{m}\right)=\frac{n_{m}a_{m}}{n_{1}a_{1}+n_{2}a_{2}+...+n_{m}a_{m}+...+n_{N}a_{N}}$.
This generalizes to the probability density function:
\begin{equation}
w_{obs}\left(a\right)=\frac{aw\left(a\right)}{\int_{0}^{\infty}a'w\left(a'\right)da'}=\frac{a}{\mu}w\left(a\right)
\end{equation}
for observing a cycle of duration $a$. Now we may consider
a cycle of duration $a'$ spanning the window of time $[t_{1}=0,t_{2}=a']$.
If the cell is observed at an arbitrary time point within that window,
the probability the cell will have been within the age range $a+\Delta a$,
is simply $\frac{\Delta a}{a'}$ if $a'\geq a+\Delta a$ and $0$ if $a'$ is smaller. Thus the probability
of observing a cell of age $a$ at an arbitrary time during a cycle of unknown duration is:
\begin{equation}
g_{Fixed}\left(a\right)=\int_{a}^{\infty}\frac{da'}{a'}w_{obs}\left(a'\right)=\int\frac{1}{a'}\frac{a'}{\mu}w\left(a'\right)da'=\frac{1}{\mu}\int_{a}^{\infty}w\left(a'\right)da'
\end{equation}
We may note using the same method as above, that when $w\left(a\right)$
is mean scalable, this ensemble is mean scalable too: $G_{Fixed}\left(x\right)=\int_{x}^{\infty}\Omega\left(x\right)dx$.
Another simple and straightforward, but useful observation is that
the mean age of the fixed ensemble can be calculated with a single
integral. Written in the usual way, it is a bit cumbersome to calculate
directly: $\overline{a}=\int_{0}^{\infty}ag\left(a\right)da=\int_{0}^{\infty}a\left(\frac{1}{\mu}\int_{a}^{\infty}w\left(a'\right)da'\right)da$.
However, we may more easily write down an expression for the mean
age of any lineage, which will be the same as the mean age of the
entire ensemble. The mean age of each lineage is simply the average
of the mean age of each cycle (the mean age of a cycle of length $a$
is simply $\frac{a}{2}$):

\begin{equation}
\overline{a}_{Fixed}=\int_{0}^{\infty}\frac{a'}{2}\frac{a'}{\mu}w\left(a'\right)da'=\frac{1}{2\mu}\int_{0}^{\infty}a'^{2}w\left(a'\right)da'
\end{equation}
In the case where $w\left(a\right)$ is gamma distributed see Eq.~\ref{eq:gammadistribution}, this yields
a closed form expression:

\begin{equation}
\overline{a}_{Fixed}=\frac{\alpha}{2\beta}\left[1+\frac{1}{\alpha}\right]=\frac{\mu}{2}\left[1+CV^{2}\right]
\end{equation}
We see that the mean age remains close to $\frac{\mu}{2}$ until the
$CV$ gets quite large (since the $CV$ is rarely greater than 1 for experimentally observed $w\left(a\right)$\cite{hashimoto2016noise,cerulus2016noise,wang2010robust,rochman2016grow}).
We may also note that in the delta function limit for the cell cycle
duration distribution, $g\left(a\right)_{Fixed}=\frac{1}{\mu}\int_{a}^{\infty}\delta\left(\mu-a'\right)da'=\frac{1}{\mu}$
is simply uniform. In contrast for the exponentially growing case,
in the limit where the $CV$ tends to zero and $w\left(a\right)$
is a delta function at $a=\mu$, we retrieve $g\left(a\right)=\frac{2\ln\left(2\right)}{\mu}\exp\left(-\frac{\ln\left(2\right)}{\mu}a\right)$.
We see in this case, the mean age is:

\begin{equation}
\overline{a}=\int_{0}^{\infty}a\frac{2\ln\left(2\right)}{\mu}\exp\left(-\frac{\ln\left(2\right)}{\mu}a\right)da=\mu\left(\frac{1}{\ln\left(2\right)}-1\right)
\end{equation}
This is about $0.44\mu$. In Fig.~\ref{fig:FixedEnsembleCartoon} we show that the fixed population
is older than the exponentially growing population when $w\left(a\right)$
is gamma distributed for all $CV$. We also want to emphasize this
is true in general, independent of the form of $w\left(a\right)$.
This can be observed from a comparison of the age distributions: $g(a)=2b\exp\left(-ba\right)\left[\int_{a}^{\infty}w(a')da'\right]$
and $g_{Fixed}\left(a\right)=\frac{1}{\mu}\int_{a}^{\infty}w\left(a'\right)da'$.
The two expressions differ only by the factor $2\mu b\exp\left(-ba\right)$
which monotonically decreases with age. This means if you examine a
snapshot of cells which have been maintained in a population of fixed
number (e.g. mother cells in a microfluidic device where only one of two daughter cells remains in the ensemble under observation after cell division), the ensemble
will be older than a group of unconstrained cells. Stated another way, if you
were to pick two arbitrary cells: one from a population of fixed number and one
from an exponentially growing population, the cell from the population of fixed number
would probably be the older cell. 
In Fig.~\ref{fig:FixedEnsembleCartoon}, we examine
how $g_{Fixed}\left(a\right)$ changes with respect to changing $CV$
of the cell cycle duration distribution, assuming gamma-distributed
$w\left(a\right)$, and compare $g\left(a\right)$ with $g_{Fixed}\left(a\right)$.
We see that there is a stronger dependence on the $CV$ of $w\left(a\right)$
for the fixed distributions. 

\section{ergodicity}

The condition, $P\left(a_{n}\rightarrow a_{n+1}\right)=w\left(a_{n+1}\right)\equiv w\left(a\right)$,
which led to the nice properties of the age distributions discussed
above is really a statement about the ergodicity of the cell
cycle duration distribution which may not be true for
the biological system of interest. The following definition is usually used
to describe an ergodic system: consider an ensemble of measurements
$x\left(t,y\right)$ made at time $t$ of individual $y$. This ensemble
is considered to be ergodic if $P\left(x\left(t_{i},y_{i}\right)\right)=P\left(x\left(t_{j},y_{j}\right)\right)$;
that is, the probability of observing state $x$ is the same at anytime
and from any individual. To clarify, this work focuses on the
case where the state observed is the cell cycle duration. Alternatively, this can be written:

\begin{equation}
\label{eq:ergoB4t1}
P\left(x\left(t_{i},y_{i}\right)\right)=P\left(x\right)
\end{equation}
where $P\left(x\right)$ is the probability of observing state $x$
conserved across all individuals at all times. We want to illustrate,
for some ensembles, that this condition implies over time the same
behavior is recapitulated as that over space. Let us consider the
case where the number of individuals does not change and without loss of generality
consider the case with a single individual. Here we are considering
situations as described in Section II, for example, within micro-fluidic devices
where after each cell division, one daughter cell is removed from the ensemble under observation. Now we may consider
the probability of making an observation within the interval $I_{x}=[x-\xi,x+\xi]$: $P\left(I_{x}\right)=\int_{x-\xi}^{x+\xi}P\left(x\right)dx$.
Let us make $n$ observations on a single individual, and calculate the measured probability
density for the interval $I_{x}$: $\frac{k}{2\xi n}$, where $k$
is the number of observations which fell within $I_{x}$. These calculated
probability densities follow the binomial distribution: $P_{n}(\frac{k}{2\xi n},n)=\left(\begin{array}{c}
n\\
k
\end{array}\right)P(I_{x})^{k}\left[1-P(I_{x})\right]^{n-k}$. The root square difference between $P_{n}\left(\frac{k}{2\xi n},n\right)$
and the mean, $\frac{P\left(I_{x}\right)}{2\xi}$, is $\frac{1}{2\xi n}\sqrt{nP\left(I_{x}\right)\left(1-P\left(I_{x}\right)\right)}\leq\frac{1}{4\xi\sqrt{n}}$.

We may readily construct the distribution $P_{n,\xi}\left(x\right)$
defined within each interval as above. For suitably well behaved $P\left(x\right)$
for which $\frac{P\left(I_{c}\right)}{2\xi}\rightarrow P(c)$ as $\xi\rightarrow0$,
we may conclude that for any $\epsilon$ there exists a $\xi$ such
that the root square difference between $P_{n,\xi}\left(x\right)$ and $P\left(x\right)$
tends to some limit below $\epsilon$ as $n\rightarrow\infty$. Smoothness
of $P\left(x\right)$ is more than sufficient and general enough for
our discussion, so we will drop the $\xi$ notation and conclude that
for smooth $P\left(x\right)$:
\begin{equation}
\label{eq:ergoB4t2}
P_{n}\left(x\right)\rightarrow P\left(x\right)
\end{equation}
as $n\rightarrow\infty$.

Thus we see that the ensemble behaves the same way over time and space: if we observe a single individual at many times,
it recapitulates the behavior of the entire ensemble at a single time. We have shown
that Eq.~\ref{eq:ergoB4t2} is implied by Eq.~\ref{eq:ergoB4t1}; however, we have only shown this
is true for the case where the number of individuals in the ensemble
does not vary. We will show that in general, neither Eq.~\ref{eq:ergoB4t2} nor Eq.~\ref{eq:ergoB4t1} is implied by the other
and will refer to them as follows:

\subsection*{Ergodicity Type I}

\emph{We will define an ensemble to be of ergodicity type I if:}

\emph{
\begin{equation}
\label{eq:conditionOfInterest}
P\left(x\left(t_{i},y_{i}\right)\right)=P\left(x\left(t_{j},y_{j}\right)\right)=P\left(x\right)
\end{equation}
In other words, the ensemble has an unbiased selection of states which
does not depend on time or the individual observed. Our condition
of interest, $P\left(a_{n}\rightarrow a_{n+1}\right)=w\left(a_{n+1}\right)\equiv w\left(a\right)$
is equivalent to Eq.~\ref{eq:conditionOfInterest}}

\subsection*{Ergodicity Type II}

\emph{We will define an ensemble to be of ergodicity type II if:}

\emph{
\begin{equation}
P_{n}\left(x\right)\rightarrow P\left(x\right)
\end{equation}
as $n\rightarrow\infty$ where $P_{n}\left(x\right)$, defined above, is the estimation of $P\left(x\right)$ made from $n$ observations of any single individual.
In other words, each individual
in the ensemble will, over time, recapitulate the behavior of the
entire ensemble.}

\begin{figure}
\includegraphics[scale=0.41]{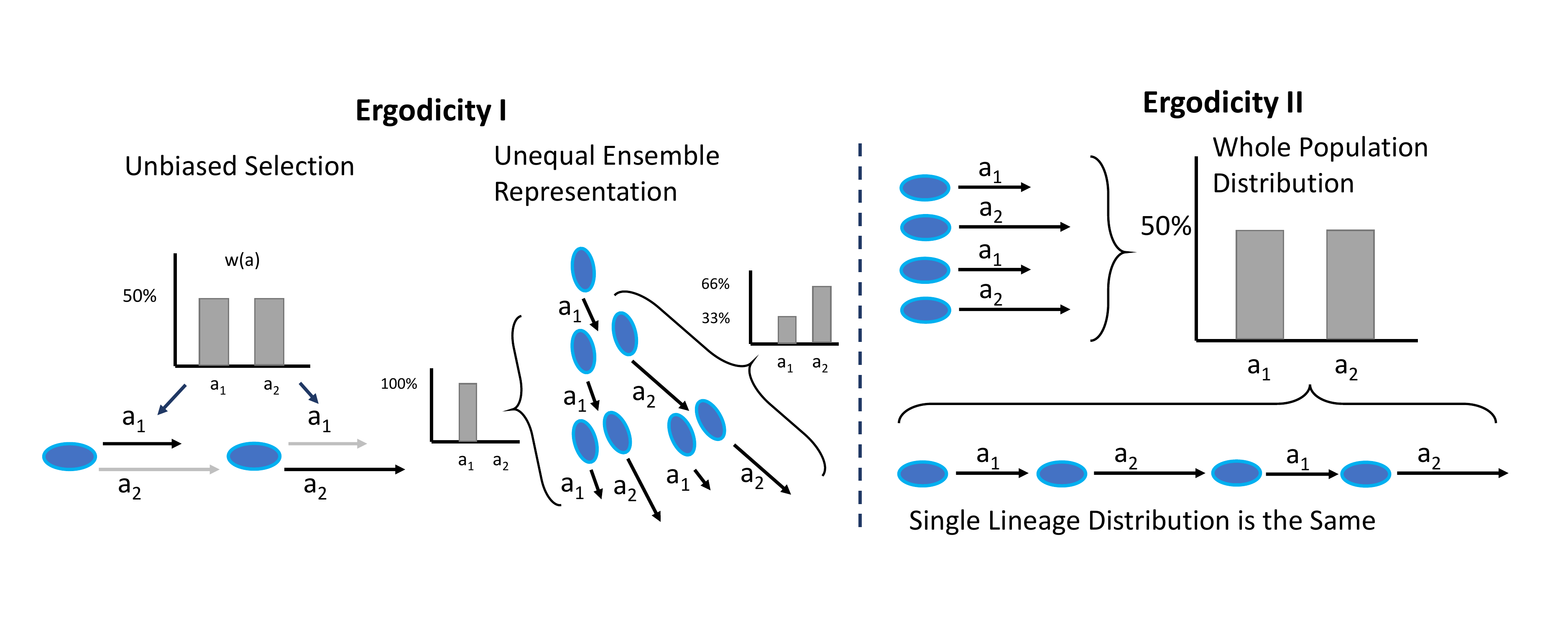}

\caption{Illustration of the two types of ergodicity: ergodicity I where the
ensemble is derived from an unbiased selection of cell cycle durations which does not depend
on time or the individual observed and ergodicity II where each individual
in the ensemble will, over time, recapitulate the behavior of the
entire ensemble.}
\label{fig:ergodicityCartoon}	
\end{figure}

We have just seen when the number of individuals is fixed, ergodicity
type I implies ergodicity type II; however the converse is not true.
Consider an ensemble composed of two individuals $x_{1}$ and $x_{2}$
which may occupy two possible states $a$ and $b$. Suppose that at
odd observations $x_{1}$ occupies $a$ and $x_{2}$ occupies $b$
and vice versa. The probability of observing either state in the entire
ensemble is $\frac{1}{2}$ at any time and $P_{n}\left(x\right)\rightarrow P\left(x\right)$
for both individuals; however $P\left(x_{1}\left(t_{i},\left[a,b\right]\right)\right)\neq P\left(x_{2}\left(t_{j},\left[a,b\right]\right)\right)$
for all times. Thus this ensemble is ergodic in the second sense but
not ergodic in the first sense. See Fig. 3 for a cartoon illustrating the
two types of ergodicity discussed.

When the number of individuals varies, as it does in the exponentially
growing ensemble, ergodicity I does not imply ergodicity II. Let us
turn to our ensemble of interest, the cell cycle duration distribution.
We may utilize the von Foerster equation as we did earlier and note
that using the same formalism assumes ergodicity I through the condition
\emph{$P\left(a_{n}\rightarrow a_{n+1}\right)=w\left(a_{n+1}\right)\equiv w\left(a\right)$}
as we discuss above. We may return to the expression for the growth rate,
$b$: $1=2\int_{0}^{\infty}\exp(-b a)w(a)da$ and note that $\exp(-b a)$
is a convex function. By Jensen's inequality this implies $E\left(f\left(a\right)\right)=\int_{0}^{\infty}\exp(-b a)w(a)da\geq \exp\left(-b\int_{0}^{\infty}aw\left(a\right)da\right)=f\left(E\left(a\right)\right)$.
Further, $f\equiv \exp(-b a)$ is strictly convex, which means equality
holds only when $a\equiv C$. Considering
the case where $w\left(a\right)=\delta\left(a-\mu\right)$:

\begin{equation}
1=2\int_{0}^{\infty}\exp(-b a)\delta\left(a-\mu\right)d\tau\Rightarrow b=\frac{\ln\left(2\right)}{\mu}
\end{equation}
Considering the case where $w\left(a\right)$ is not a Dirac delta function
but a distribution with the same mean:

\begin{equation}
1=2\int_{0}^{\infty}\exp(-b a)w(a)da>2\exp(-b\mu)\Rightarrow b>\frac{\ln\left(2\right)}{\mu}
\end{equation}
Thus, for any non-delta function distribution, $b>\frac{\ln\left(2\right)}{\mu}$.
This means that $P_{n}\left(x\right)$ does not tend to $P\left(x\right)$
as the mean of $P_{n}\left(x\right)$ tends to a value below the mean of $P\left(x\right)$. Thus ergodicity
type I does not imply ergodicity type II when the number of individuals
varies.  In fact, for this example, ergodicity type I and ergodicity type II are mutually exclusive.
This brings us to a statement of the relative strengths of these conditions (Please note that for the case of the exponentially growing ensemble discussed above,
many observations of a single individual refers to a lineage of cell cycle durations,
for example, current duration, mother cell duration, grandmother cell duration,...):

\subsection*{Hierarchy of Ergodicities Types I and II}

\emph{For an ensemble with a fixed number of individuals, ergodicity
type I implies ergodicity type II; however, ergodicity type II does
not not imply ergodicity type I. Thus for these ensembles, ergodicity
type I is the stronger condition. For an ensemble with a variable
number of individuals, being of ergodicity type I does not necessarily
imply the ensemble is of ergodicity type II and vice versa. Furthermore, for some
ensembles, including the cell cycle duration distribution discussed above, they are
mutually exclusive conditions.
This distinction is of interest relative to the qualitative notion
that an ergodic ensemble,}\emph{\noun{ }}\textsc{\small{}recapitulates
within a single individual over many observations the behavior of
the entire ensemble at a single observation}\emph{, or }\textsc{\small{}behaves
the same way over time and space}\emph{. These statements represent
ergodicity type II and may not accurately describe an ensemble of
ergodicity type I with a varying number of individuals.}

\section{Growth Rate Gain}

We have discussed how the cell cycle duration distribution is of ergodicity
I under the condition \emph{$P\left(a_{n}\rightarrow a_{n+1}\right)=w\left(a_{n+1}\right)\equiv w\left(a\right)$}
which is assumed in the traditional formalism used to solve the von
Foerster equation. Thus we know unless  $w\left(a\right)\equiv\delta(a-\mu)$,
the growth rate is higher than the \qts{naive} growth rate $\frac{\ln\left(2\right)}{\mu}$
and the bulk doubling time is shorter than the mean division time.
This phenomenon is commonly referred to as the \qts{growth rate gain}.
It was discussed first as far back as the 1950's\cite{powell1956growth}
and has been the object of renewed recent interest\cite{cerulus2016noise,hashimoto2016noise}.
At the center of the issue sits the finding that when the duration
of mother-daughter cell cycles are positively correlated, the growth rate
increases and when they are negatively correlated, the growth rate
decreases. We have shown that even in the case without explicit correlation,
ergodicity type I, the growth rate gain appears. This is due to the
nature of the growing ensemble. When cells which divide quickly are equally
likely to form daughter cells which divide quickly as they are to
form daughter cells which divide slowly, this leads to an unequally
large number of short cell cycles represented in the ensemble (see
Fig.~\ref{fig:ergodicityCartoon}). On the other hand, we will show if the the ensemble is of
ergodicity II, then there is no growth rate gain: the bulk doubling
time is equal to the mean cell cycle duration. 

Consider an exponentially growing ensemble which begins with a single
individual such that all individuals in the ensemble will be of ergodicity
II. Suppose at any given time,
the ensemble contains cell cycle durations structured into $L$
lineages each of length $N_{l}$. Let us consider the \qts{error} of a single lineage,
the root square difference between the mean of the lineage and $\mu$,
to be $\frac{\epsilon_{l}}{\sqrt{N_{l}}}$. The composite error
of all the lineages within the ensemble labeling $max_{L}\left(\epsilon_{l}\right)\equiv\epsilon$,
 and $min_{L}\left(N_{l}\right)\equiv N$ is:

\begin{equation}
E=\sum_{l=1}^{L}\frac{\epsilon_{l}}{\sqrt{N_{l}}}\leq\epsilon\frac{L}{\sqrt{N}}
\end{equation}

Thus if all lineages are of ergodicity type II, the ensemble must
also be of ergodicity type II. The bulk doubling
time is is the mean of every cell cycle observed in the population
(over some window of time observed) - which if the ensemble is of
ergodicity II, is simply $\mu$.

\section{Explicit Correlation}

\begin{figure}
\includegraphics[scale=0.35]{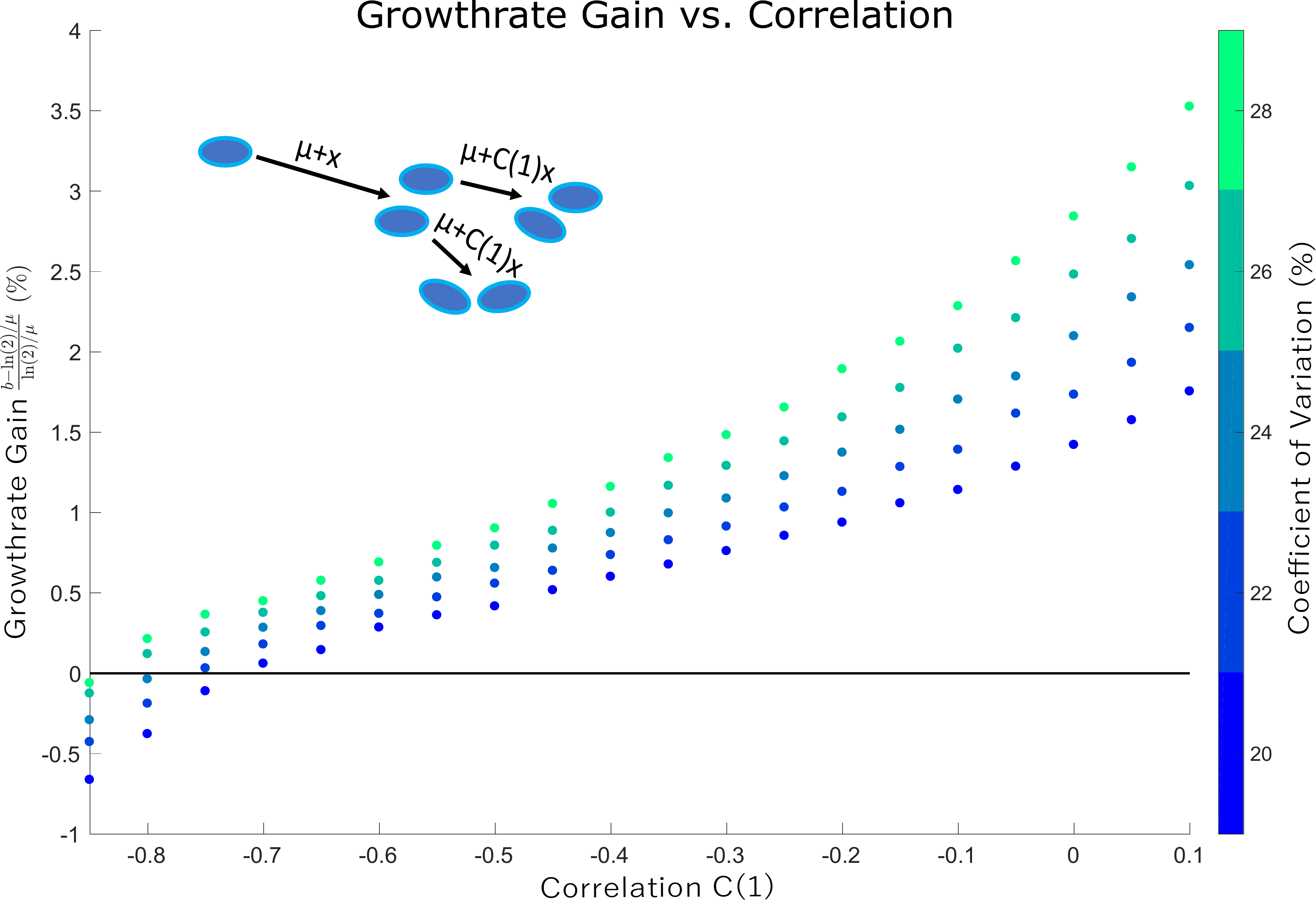}

\caption{A display of the growth rate gain as a function of ensemble $CV$ and
mother-daughter cell cycle duration correlation $C(1)$. A strong,
negative mother-daughter correlation is required to return the ensemble
to the naive bulk growth rate of $\frac{\ln\left(2\right)}{\mu}$. The
cartoon illustrates that a cell which divided after a cycle of duration
$\mu+x$ is most likely to form daughter cells which attain cycles
of duration $\mu+C\left(1\right)x$.}
\label{fig:Fig4}
\end{figure}

We have shown when ergodicity I is maintained in the case of the cell
cycle duration distribution, $P\left(a_{n}\rightarrow a_{n+1}\right)=w\left(a_{n+1}\right)\equiv w\left(a\right)$,
the ensemble growth rate is higher than $\frac{\ln\left(2\right)}{\mu}$.
This growth rate gain may be attributed to the prevalence of lineages
largely comprising cell cycles shorter than the mean. While there is
no explicit mother-daughter correlation within the selection of cell
cycle durations, there is an effective correlation arising from this
variable weighting of lineages. We sought to probe the degree of this
effective correlation through the addition of explicitly negative
mother-daughter correlation. In this way, we could find the degree
of negative correlation which must be imposed to return an ensemble
to the naive growth rate, $\frac{\ln\left(2\right)}{\mu}$. We chose
a form for $P\left(a_{n}\rightarrow a_{n+1}\right)$ based on the
model presented in\cite{rochman2016grow}:

\begin{equation}
P\left(a_{n}\rightarrow a_{n+1}\right)=A\exp\left[-\frac{1}{2\sigma_{1}^{2}}(a_{n+1}+a_{n}-2\mu)^{2}\right]\exp\left[-\frac{1}{2\sigma_{2}^{2}}(a_{n+1}-a_n)^{2}\right]
\end{equation}
The autocorrelation function generated from $P\left(a_{n}\rightarrow a_{n+1}\right)$
is: 

\begin{equation}
C(n)=\left(\frac{\sigma_{1}^{2}-\alpha\sigma_{2}^{2}}{\sigma_{1}^{2}+\sigma_{2}^{2}}\right)^{n}
\end{equation}
The object of interest here is $C\left(1\right)$, the mother-daughter
correlation. We simulated lineage trees of cell cycle durations with
each successive cycle duration drawn from $P\left(a_{n}\rightarrow a_{n+1}\right)$.
See Fig.~\ref{fig:Fig4}.
The first generation cycle duration was drawn from a Gaussian distribution
with mean $\mu$ and the standard deviation corresponding to the stationary
distribution. Without loss of generality, $\mu$ was taken to be $20$
(unitless) and $\sigma_{1}$ and $\sigma_{2}$ were chosen to obtain
the desired $C\left(1\right)$ correlation and $CV$. Each cell lineage simulation
was continued until a generation was reached where the cell cycle
duration distribution was sufficiently close to the steady-state distribution
evolving from $P\left(a_{n}\rightarrow a_{n+1}\right)$. The test
distribution was considered to be sufficiently close to the steady-state
when the relative root square difference was no more than $5\%$.
Once the steady-state was reached, the growth rate of the population
was calculated as the slope of the best-fit line to the logarithm
of average cell number over $100$ trials.

We found a very substantial degree of negative correlation ($C(1)\approx-0.75)$
must be imposed to return an ensemble to the naive growth rate, $\frac{\ln\left(2\right)}{\mu}$;
however, this phenomenon depends on the $CV$ of the cell cycle duration
distribution since the growth rate gain is higher when the $CV$ is
larger. In other words, for an ensemble to attain a bulk growth rate
of $\frac{\ln\left(2\right)}{\mu}$ given only mother-daughter correlations
(i.e. grandmother-granddaughter cell cycle durations are uncorrelated),
an anti-correlation of about $75\%$ must be imposed. Under these
conditions, a cell which divided after a cycle of duration $\mu+x$
is most likely to form daughter cells which attain cycles of duration
$\mu-0.75x$. Stated another way, this implies that ensembles which
enforce ergodicity I are essentially approximately $75\%$ correlated.

\section{Discussion}

We have seen that the concept of ergodicity bifurcates into two distinct
properties within ensembles that have a variable number of individuals:
ergodicity I, the unbiased selection of states and ergodicity II,
the recapitulation of whole-ensemble behavior from any single individual.
Many of the nice properties of the traditional formalism used to solve
the von Foerster equation rely on the assertion that the cell cycle
duration distribution is of ergodicity I. Under these conditions, the
bulk growth rate of the population is higher than $\frac{\ln\left(2\right)}{\mu}$
due to a disproportionately large number of short cell cycles represented
in the ensemble. When the ensemble is of ergodicity II, similar to
the condition where there is a negative mother-daughter cell
cycle duration correlation, the growth rate returns to $\frac{\ln\left(2\right)}{\mu}$.
This observation corroborates the idea that
the relationship between the variability of the cell cycle duration distribution
and the population growth rate is impacted by mother-daughter correlations\cite{lin2017effects}.
Since the \qts{growth rate gain} observed within ensembles of ergodicity
I stems from disparities between lineages composed of primarily short
cell cycle durations and those of long cell cycle durations, the larger
the $CV$ of the cell cycle duration distribution, $w\left(a\right)$,
the greater the effect. In the most basic sense, this growth rate gain
comes from asymmetric divisions. When ergodicity II is not enforced,
lineages of exclusively short cell cycles arise. Many sister cells
of cells within these lineages have long cell cycle durations. This
phenomenon may be purely due to stochasticity present after division
has finished or it might be due to programmatic asymmetry in the allocation
of resources to each daughter cell. The use of the ``FUCCI'' live-cell cell cycle phase
labeling system\cite{sakaue2008visualizing} will continue to provide the opportunity for experimental
validation of these concepts relating to the age structure of the population\cite{sandler2015lineage}. In particular
it may be interesting to evaluate how the age distribution of a monolayer varies as it closes a wound. In this case, one would expect
cells on the edge of the wound to assume a different age distribution as they divide more frequently or migrate faster than cells far
from the wound. Recently, related perspectives on
how ergodicity and the lineage structure impacts growth rate have brought more
interest to this topic\cite{thomas2017making,thomas2017single}.

Asymmetric division, often a complex
process\cite{mena2017asymmetric}, has been well
established in a variety of pro- and eukaryotes\cite{li2013art} including
the orchestration of stem cell differentiation and self-renewal.
It has been shown that even E. coli display complex polar protein
localization\cite{lybarger2001polarity} and that pathological polar
aggregates can be asymmetrically inherited which may increase fitness
by \qts{rejuvenating} the daughter cell that accepts less damage\cite{winkler2010quantitative}.
Completely symmetric division requires cells to fix inherited damage;
otherwise, all cells will eventually have accumulated critical amounts.
There are likely to be costs associated with coordinating asymmetric
division and inherited damage mitigation which are balanced in optimal
growth strategies. It has been reported that under some conditions,
E. coli age within ten generations\cite{stewart2005aging} while under
others, stable growth has been observed for hundreds of generations\cite{wang2010robust}.
Perhaps in the latter case, higher growth rates can be achieved through
damage mitigation in all cells than the asymmetric inheritance of
damage and subsequent loss of the damaged population.

\subsection*{Author Contributions}

NDR conducted the analysis in sections II, III, and IV. DMP completed
the simulations in section V. NDR, DMP, and SXS wrote the paper.
\begin{acknowledgments}
Thanks to Daniel Fischer for his portion of the \qts{Convexity and equality
in Jensen inequality} post on math.stackexchange.com. Nash Rochman's
work was supported by an NIH NTCR T-32 grant.
\end{acknowledgments}

\bibliographystyle{plain}
\bibliography{biblio}

\end{document}